\documentclass[sigconf]{acmart}

\usepackage{enumitem}
\usepackage{hyperref}
\usepackage{tabularx}
\usepackage{xurl}  
\usepackage{graphicx}
\usepackage{amsmath}
\usepackage{hyperref}
\usepackage{multirow, multicol, makecell}
\usepackage{pifont}
\usepackage{mdframed}
\usepackage{framed}

\newcommand{\Name}{\textsc{DePro}}
\newcommand{\ignore}[1]{}


\title{\textsc{DePro}: Understanding the Role of LLMs in Debugging Competitive Programming Code}

\author{Nabiha Parvez}
\affiliation{%
    \institution{Military Institute of Science And Technology}
    \city{Dhaka}
    \country{Bangladesh}
    }
    \email{nabihaparvez11@gmail.com}
\author{Tanvin Sarkar Pallab}
\affiliation{%
    \institution{Military Institute of Science And Technology}
    \city{Dhaka}
    \country{Bangladesh}
    }
    \email{tanvin.pallab2442002@gmail.com}

\author{Mia Mohammad Imran}
\affiliation{%
  \institution{Missouri University of Science and Technology}
  \city{Rolla, Missouri}
  \country{USA}
}
\email{imranm@mst.edu}

\author{Tarannum Shaila Zaman}
\affiliation{%
  \institution{University of Maryland Baltimore County}
  \city{Baltimore, Maryland}
  \country{USA}
}
\email{zamant@umbc.edu}

\begin{document}

\begin{abstract}
Debugging consumes a substantial portion of the software development lifecycle, yet the effectiveness of Large Language Models (LLMs) in this task is not well understood. Competitive programming offers a rich benchmark for such evaluation, given its diverse problem domains and strict efficiency requirements. We present an empirical study of LLM-based debugging on competitive programming problems and introduce {\Name{}}, a test-case–driven approach that assists programmers by correcting existing code rather than generating new solutions. \Name{} combines brute-force reference generation, stress testing, and iterative LLM-guided refinement to identify and resolve errors efficiently. 
Experiments on 13 faulty user submissions from \textit{Codeforces} demonstrate that \Name{} consistently produces correct solutions, reducing debugging attempts by up to 64\% and debugging time by an average of 7.6 minutes per problem compared to human programmers and zero-shot LLM debugging.
\end{abstract}

\acmBooktitle{Companion Proceedings of the 34th ACM Symposium on the Foundations of Software Engineering (FSE '26), June 5--9, 2026, Montreal, Canada}
\maketitle

\section{Introduction}
Debugging, the process of identifying and fixing software defects, is a multi-step procedure involving bug identification, localization, reproduction, root cause analysis, and resolution~\cite{def:debug}. It is estimated that debugging consumes nearly 50$\%$ of the software development life cycle \cite{Steve:Code}. Effective debugging requires not only familiarity with the source code but also reasoning about root causes; incomplete or superficial fixes often introduce additional errors~\cite{Lu:2008:LMC:1353535.1346323}.

Large Language Models (LLMs)~\cite{10903889} have recently transformed Natural Language Processing (NLP)~\cite{lota5060080recent} and code-related tasks such as algorithmic problem solving and code generation~\cite{chen2021evaluatinglargelanguagemodels, 11028406}. 
Their growing adoption in software engineering raises an important question: \textit{to what extent can LLMs assist in debugging, particularly in complex, real-world scenarios?} 
Despite progress in LLM-based development tools (e.g., Copilot~\cite{copilot}, AlphaCode~\cite{Alphacode}), systematic evaluation of their debugging capabilities remains limited.

Competitive programming provides a rich and rigorous testbed for evaluating debugging techniques. These problems span diverse domains, impose strict computational constraints, and demand both efficiency and correctness, making them well-suited for assessing the debugging capabilities of LLMs~\cite{11028406}.

Recent work has analyzed LLMs’ ability to solve competitive programming problems~\cite{11028406}. Other studies have investigated LLM-assisted debugging in broader contexts, including interactive debugging~\cite{10.1145/3729355}, scientific debugging frameworks~\cite{kang2023explainableautomateddebugginglarge}, and multi-agent collaboration for code repair~\cite{ashrafi2025enhancingllmcodegeneration}. 
However, research on debugging within competitive programming remains limited, and systematic evaluations of LLM performance in this context are lacking. 
Given the diversity and rigor of competitive programming challenges, analyzing LLMs’ zero-shot debugging effectiveness offers a meaningful measure of their overall debugging capabilities.

In this paper, we first present an empirical and manual study on 10 different problems from Codeforces \cite{codef}. For each problem, we selected 10 users who had repeatedly received incorrect answers. For each user, we took their faulty code and asked ChatGPT-5 \cite{gpt5} to debug it—not by generating new code, but by correcting the existing code. We then analyzed the results and found that the LLM generally required fewer attempts than human users. However, in some cases, the LLM needed many attempts; for instance, it took 8 attempts to fix a single user’s fault. We manually investigated the causes behind these high-attempt cases and also observed differences in the LLM’s code-solving strategies. Our manual study summary is as follows: LLMs are effective at reproducing standard solution patterns and applying localized fixes, but they lack the flexibility and strategic reasoning that human programmers employ, particularly for problems with multiple valid solutions or unconventional approaches.

We propose an automated approach, \Name{}, based on the observation that LLMs debug more effectively when provided with specific problem details, particularly failing test cases. \Name{} first generates a brute-force reference solution using ChatGPT-5, then stress-tests both the user’s code and the reference on random and edge-case inputs to identify failures. The LLM is iteratively prompted with the failing input, expected output, and user code to suggest fixes, with each modification retested for up to eight iterations. This process enables the LLM to refine code, identify errors, and improve debugging efficiency.

We evaluate \Name{} on 13 cases drawn from 6 faulty Codeforces problems \cite{codef}, each submitted by a different user, and find that it can effectively debug all 13 problems, producing correct solutions in an average of 6.5 minutes per problem.

\subsection{Empirical Study}
\subsubsection{Data Collection} 
We curate a dataset of 10 problems, each with 10 user codes, resulting in a total of 100 user codes. We take these codes from the popular competitive programming platform Codeforces \cite{codef}. 
We select the problems based on user attempts, focusing on cases where users submit multiple wrong answers, struggle to debug their solutions, but ultimately solve the problems after several tries. 
An attempt is defined as a single modified submission intended to correct the previous error.
The problem set includes constructive algorithms (4), math (2), greedy (3) and implementation (1). Their difficulty ranges from easy (3) to intermediate (7).

We also select user codes based on the number of attempts required to solve the problem. In particular, we include codes that fail on the first submission. To ensure variety, we choose codes with different failure points across test cases (e.g., TC-2, TC-8, TC-15) rather than selecting codes that fail at the same point. This approach prevents the dataset from biasing toward a single debugging pattern. We test GPT-5’s reasoning model by providing the LLM with the problem statement, time complexity, memory complexity, input, and output before giving it the user’s code. We then use Prompt 1 to extract the debugging feedback generated by the LLM.

\ignore{
\begin{center}
\small 
\begin{tabular}{|p{0.28\textwidth}|p{0.25\textwidth}|}
\hline
\textbf{Problem Statement} &
There is a binary string \(a\) of length \(n\). In one operation.. \\
\hline
\textbf{Time Limit} & per test 1 second \\
\hline
\textbf{Memory Limit} & per test 256 megabytes \\
\hline
\textbf{Input} &
The first line contains a single integer \(t\) \((1 \le t \le 10^4)\) — the number of test cases.... \\
\hline
\textbf{Output} &
For each test case, output "YES" if it is... \\
\hline
\textbf{Initial Prompt} &
This is the problem. Can you generate a brute-force solution? \\
\hline

\end{tabular}
\normalsize
\end{center}
}

\begin{promptbox}{Prompt 1: Debugging Feedback Extractor Prompt}
{
\small
\noindent
\textbf{Problem Statement:} You are given an array a1,a2,...
\newline
\textbf{Time Limit:} 1 second  \hspace{0.2cm} \textbf{Memory Limit:} 256 megabytes
\newline
\textbf{Input}:
Each test contains multiple test cases... 
\newline
\textbf{Output}: 
For each test case, output a single integer...
\newline
\textbf{Sample Input:} 8 15 22 30
\hspace{0.2cm}
\textbf{Sample Output:} 7 \newline
\textbf{Task}:Consider the given problem description. Here is my code for this problem. I got wrong answer. Can you debug my code? 
}
\end{promptbox}\label{prompt}

\subsubsection{Data Preprocessing:} We remove unnecessary spacing and newlines, then decompose each problem into distinct components, such as, problem statement, time constraint, memory constraint, input, output, and initial prompt to improve clarity. We clarify and simplify mathematical notations before presenting the problems to the LLM.

\subsubsection{Empirical Study Result:} 
We analyze the LLM’s responses after administering Prompt 1. Our primary objective is to compare the number of attempts the LLM requires to solve a problem with the number of attempts made by human problem solvers. The first three columns of Table \ref{studytable} present the problem name, difficulty level, and problem type. Columns four and five report the average number of attempts made by the problem solvers and by the LLM, respectively. For each problem, we collect data from ten problem solvers. On average, human problem solvers require 3.68 attempts to solve a problem, whereas the LLM requires 1.8 attempts.



\begin{table}[tb]
\caption{Summary of the empirical study on LLM performance} 
\label{studytable} 
\scalebox{.7}{
\begin{tabular}{|l|l|l|r|r|r|}
\hline
\textbf{Name} & \textbf{Dif} & \textbf{Type} & \textbf{N\_U} & \textbf{N\_L}  & \textbf{Sim.} \\
\hline
Flip Bits & Intrm & Con. Algo. & 4.2 & 1.3 & 3.8 \\
\hline
Equal with mod & Intrm & Con. Algo. & 6.7 & 4.2 & 1.9 \\
\hline
Alternating & Easy & Implement & 4 & 2 & 4.4 \\
\hline
Stable Groups & Intrm & Implement & 3.8 & 1 & 5\\
\hline
Good Start & Intrm & Implement & 3.7 & 2.6 & 4.3 \\
\hline
Cherry Bomb & Intrm & Implement & 4 & 1.3 & 3.8 \\
\hline
T-primes & Intrm & Implement & 3.2 & 1.2 & 4.7 \\
\hline
Bobritto Banditto & Intrm & Implement & 2.6 & 1.9 & 2.7 \\
\hline
Letter Home & Intrm & Implement & 2.2 & 1.5 &  4\\
\hline
Brightness Begins & Intrm & Implement & 2.4 & 1 & 3 \\
\hline
\end{tabular}
}
\begin{flushleft}
\small{
$\#Dif$ = Difficulty. $N\_U$ = Average number of manual attempts by the user to solve a problem. $N\_L.$ = Average number of attempts by the LLM to solve a problem. $Sim.$= Similarity Score. }
\end{flushleft}
\end{table}

\subsection{Manual Study}
Two competitive programmers analyze our dataset and compare the problem-solving approaches of human participants and the LLM. For the manual study, they first construct a table similar to Table \ref{manualtable}. Drawing on their experience, they describe the human and LLM debugging approaches, which appear in Columns 3 and 4 of Table \ref{manualtable}. They then assign a similarity score from 1 to 5 to indicate how closely the two approaches align and record it in the final column. We compute the similarity score by averaging the two programmers’ ratings. The final column of Table \ref{studytable} reports the average similarity score for each problem. Through this manual study, we address three research questions.
\begin{table*}[tb]
\begin{center}
\caption{Summary of the manual study on LLM performance} 
\label{manualtable} 
\scalebox{.75}{
\begin{tabular}{|l|l|p{8cm}|p{7.5cm}|r|}
\hline
\textbf{Name} & \textbf{User} & \textbf{Approach (User)} & \textbf{Approach (LLM)} & \textbf{Sim.} \\
\hline
Cherry Bomb & u1 &
Collect known pairs $b_i\ge0$, form $s_i=a_i+b_i$. If multiple distinct $s_i$ → 0; if one $x$ verify $0\le x-a_i\le k$ for missing $b_i$; if none count integers in $[\max_i a_i,\ \min_i(a_i+k)]$. &
Enforce single forced $x$ (or none), validate only missing-$b_i$ indices, and use count $=\max(0,\min_i(a_i+k)-\max_i a_i+1)$. & 5 \\
\cline{2-5}

& u2 &
Set first forced $x=a_i+b_i$ but later overwrote it; used global min/max for all-$-1$ case. &
Keep-first-$x$ (or second-pass verify all known sums); clamp negative counts to $0$. & 5 \\
\hline
Bobritto Bandito & u1 &
Shrank only right by $(n-m)$ (assumes unilateral growth), may exclude $0$. &
Split rewind: $\text{take\_left}=\min(n-m,-l)$, $\text{take\_right}=(n-m)-\text{take\_left}$; $l'=l+\text{take\_left}$, $r'=r-\text{take\_right}$. & 3 \\
\hline
\end{tabular}
}
\end{center}
\end{table*}

\noindent
\textit{\textbf{RQ1:} Is the LLM’s response relevant to the user’s solution? }

When humans and the LLM converge on similar approaches, the resulting solutions typically follow canonical or well-known strategies, indicating that the LLM effectively reproduces documented solution patterns. For purely mathematical problems, the LLM often uses different formulas than human solvers. In contrast, when humans rely on unconventional heuristics, shortcuts, or domain-specific tricks, the LLM tends to diverge, producing alternative solutions that are syntactically correct but sometimes less efficient, more verbose, or structurally different. We also observe that the LLM struggles with problems that admit multiple valid solutions, as it usually converges on a single reasoning template rather than exploring diverse approaches.


\noindent
\textit{\textbf{RQ2:} How do the LLM’s answers differ from users’ code in terms of approach?}

The LLM’s debugging behavior differs from humans’. Humans locate errors by testing boundary cases, reasoning backward, or restructuring their approach. The LLM tends to make localized edits around failing test cases, adjusting loops, conditions, or data structures based on learned patterns. While this often fixes issues quickly, it can repeat ineffective fixes or reintroduce past errors. This highlights a key distinction: humans rely on flexible reasoning and strategic exploration, whereas the LLM primarily uses pattern-based refinements.
\\
\noindent
\textit{\textbf{RQ3:} How can we improve the performance of LLM?}

We evaluate an LLM’s performance by its ability to transform buggy code into an accepted solution when provided with failing test cases. The LLM is first prompted to debug the user’s code (Prompt 1), and the generated solution is submitted to Codeforces. If the submission fails, the corresponding failing test case is provided along with the code. We find that supplying failing test cases significantly improves debugging performance compared to re-prompting without additional information. This observation motivates \Name{}, which leverages failing test cases to enhance LLM-assisted debugging.

\section{Methodology and Evaluation} Based on our manual observations, we find that LLMs perform more effectively in debugging when prompted with specific details of the problem, particularly the failing test case. Motivated by this finding, we develop an automated technique, \Name{}, that leverages LLM-assisted debugging and evaluate it on 13 real-world problems.

\subsection{{\Name}  Architecture:} 
\begin{figure}[htb]
\centering
\includegraphics[scale=0.3]{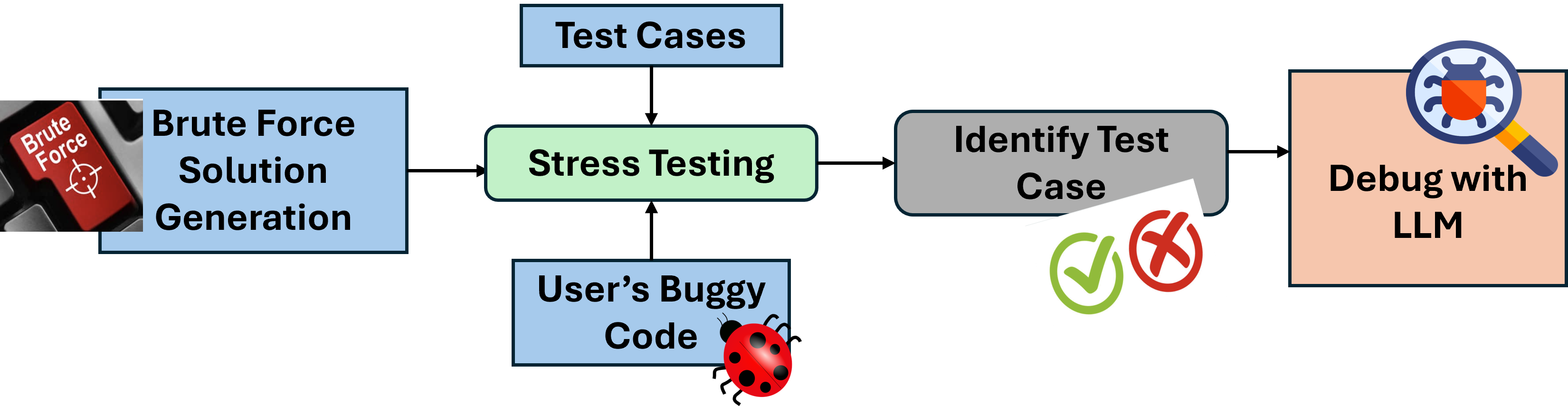}
\caption{Overview of \Name{} Methodology}
\label{fig:overview}
\end{figure}

Figure \ref{fig:overview} illustrates the workflow of \Name{}, which consists of three major steps: (1) initial code generation, (2) stress testing and identification of the failure-inducing test case, and (3) debugging with the LLM.

\noindent
\textbf{Initial Code Generation:} 
The process begins by prompting ChatGPT-5’s reasoning model with a problem template that we construct, which includes the problem description, constraints, and input-output specifications. We instruct the LLM to generate a brute-force solution that prioritizes passing all sample test cases. This solution serves as a reference for subsequent debugging steps and provides a benchmark to compare against the user’s original code.\\
\textbf{Stress Testing and Failure Detection:} After generating the brute-force solution we conduct stress testing with the user's original code. We execute both programs on a wide range of generated test cases, including edge cases and boundary values. A separate script generates the inputs, with ranges according to the problem statement. The test case generation helps to identify any mismatch between the brute-force solution and the user code. The stress testing loop terminates when it encounters a failing test case. For each failing case, we extract and store both the input and the resulting output for use in the subsequent debugging phase.\\
\textbf{Iterative Debugging with LLM:} After identifying a failing test case, we prompt the LLM with the failing input, the expected output, and the user’s original code. The model is instructed to analyze and debug the code, as illustrated in Prompt 2. Once the LLM suggests modifications, we subject the updated code to another round of stress testing. This process creates a feedback loop in which debugging is guided primarily by failing test cases. The loop of code modification and stress testing continues for up to eight iterations. In each iteration, the model refines the code’s logic, learns from previous failures, and identifies potential sources of error.

\begin{promptbox}{Prompt 2: \Name{} Debugging Prompt}
{
\small
\noindent
This is my code. It failed in the test case: \newline 
Input: 1 2 3 4 \hspace{0.2cm} Output: 2 \newline 
Expected Output: 4 \newline 
Can you debug my code? 
}
\end{promptbox}\label{prompt}



\begin{figure}[tb]
\centering
\includegraphics[scale=0.16]{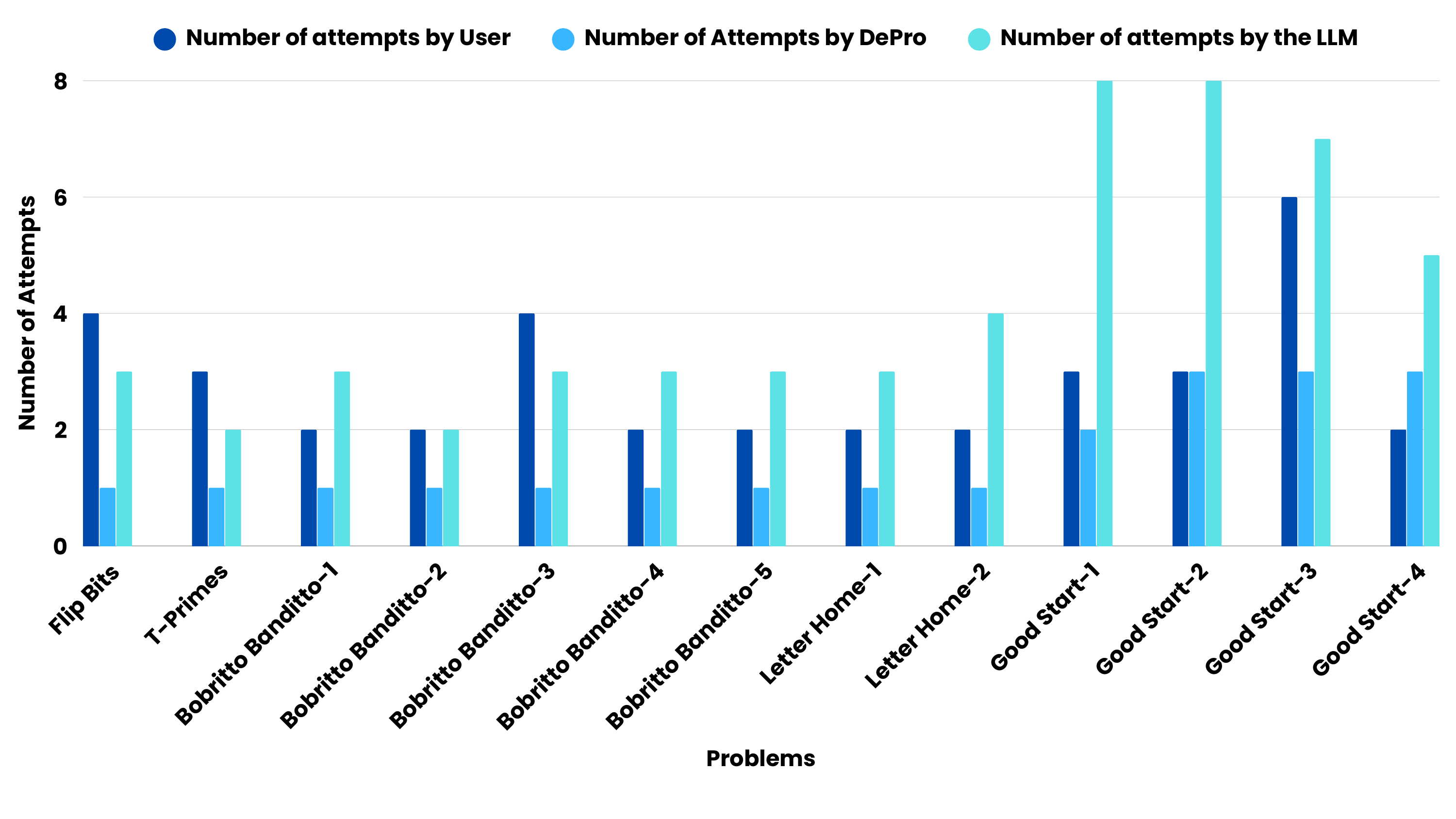}
\caption{Attempts required by \Name{}, manual methods, and zero-shot LLMs}
\label{fig:rQ1}
\end{figure}

\subsection{Evaluation}
We select 13 faulty user codes from 6 problems, for which the LLM requires an average of 4 attempts to debug, with a maximum of 8 attempts and a minimum of 2 attempts. We evaluate the effectiveness and efficiency of \Name{}. 


We implemented a prototype of \Name{} using ChatGPT-5’s reasoning model~\cite{gpt5}. The prototype consisted of three main C++ files: (i) a brute-force solution, (ii) a user-provided solution, and (iii) a random test generator. These components were orchestrated and executed together via a batch script~\cite{Batch}. During execution, if the output of the user’s solution differed from that of the brute-force solution, the corresponding failing test case was automatically displayed in the terminal. We conducted the experiments on a single workstation with an 11th-Gen Intel® Core™ i7-1165G7 @ 2.80 GHz, 16 GB RAM, a graphics device with 2 GB VRAM.



Figure \ref{fig:rQ1} presents a comparison of the total number of attempts required by human problem solvers, zero-shot LLMs, and \Name{} for each problem. On average, \Name{} reduces the number of attempts by 64\% compared to the zero-shot LLM approach described in Section 2, with a maximum reduction of 75\% and a minimum of 50\%. Across all 13 faulty user codes, \Name{} consistently requires fewer attempts than both human solvers and zero-shot LLMs, successfully producing the correct solution after debugging the user’s code. These results demonstrate that \Name{} is highly effective at debugging a variety of problem types.

\begin{figure}[tb]
    \centering
    \includegraphics[width=0.95\columnwidth]{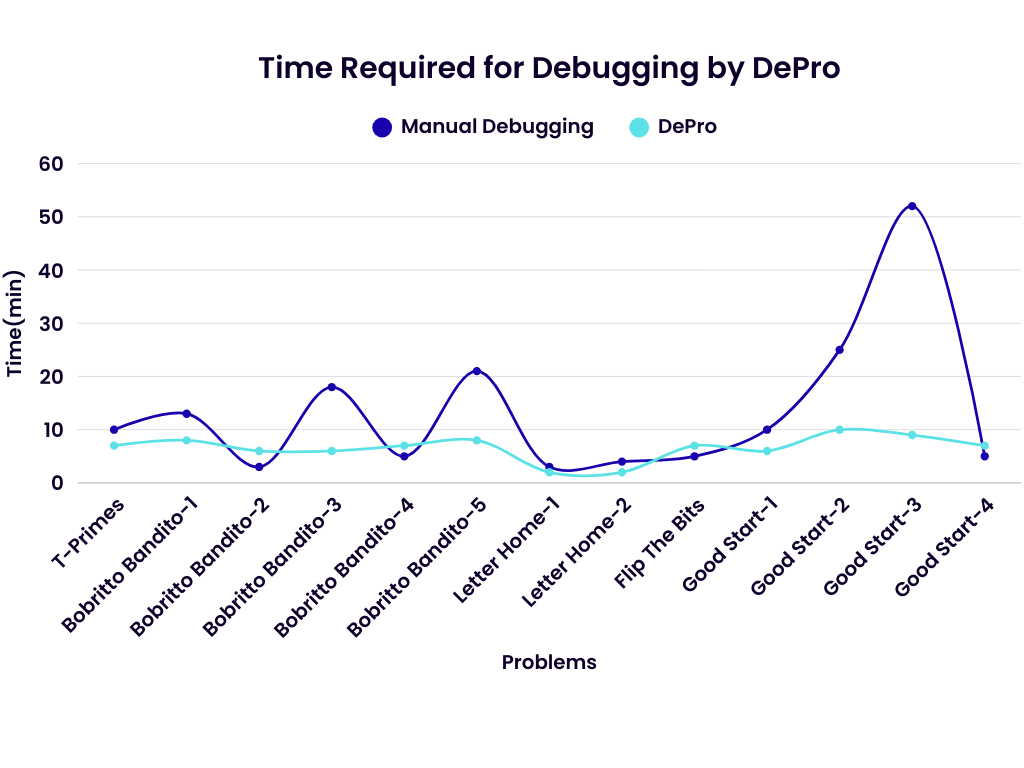}
    \caption{Total time in Debugging}
    \label{fig:Rq2}
\end{figure}

\Name{} reduces the overall time required to debug competitive programming problems, taking an average of 6.5 minutes per problem. \textbf{Figure \ref{fig:Rq2}} compares the debugging times of \Name{} and human problem solvers across our 13 user codes. On average, \Name{} reduces the debugging time by 7.6 minutes per problem compared to manual methods. These results demonstrate that \Name{} is highly efficient in debugging competitive programming solutions.

\section{Related Work}
LLM-based debugging has gained traction, with studies exploring strategies from agent-based collaboration to runtime feedback.

\noindent
\textbf{Interactive Debugging.} Systems have been developed that integrate LLMs with debuggers for natural‐language interaction. ChatDBG~\cite{10.1145/3729355} connects LLMs to GDB/LLDB, allowing queries about program state. AutoSD~\cite{kang2023explainableautomateddebugginglarge} combines LLMs with hypothesis generation and testing, and DebugBench~\cite{tian2024debugbench} introduces a benchmark for evaluating LLM debugging across multiple error types.  

\noindent
\textbf{Runtime Feedback.} Several approaches incorporate execution feedback into the debugging process. LDB~\cite{ashrafi2025enhancingllmcodegeneration} uses runtime traces to validate program blocks. Ledex~\cite{jiang2024ledex} applies execution-guided refinement to train LLMs to self-debug and provide explanations. RTLFixer~\cite{tsai2024rtlfixer} addresses hardware description languages by using compilation feedback to repair RTL syntax errors.  

\noindent
\textbf{Agent-Based Debugging.} Multi-agent frameworks assign different roles to the debugging process. FixAgent~\cite{lee2024unifieddebuggingapproachllmbased} separates tasks such as fault localization and patch generation. RGD~\cite{jin2024rgd} and COAST~\cite{yang2025coast} explore role-based collaboration and data synthesis. AGDebugger~\cite{epperson2025agdebugger} introduces user-driven steering mechanisms for multi-agent debugging systems. 

\noindent
\textbf{Automated Program Repair.} Recent advances in LLM-based automated program repair have shown promising results in fixing real-world bugs. AutoCodeRover~\cite{zhang2024autocoderoverautonomousprogramimprovement} combines LLMs with program structure-aware code search to automatically resolve Github issues, achieving 19\% efficacy on SWE-bench. SWE-agent ~\cite{yang2024sweagentagentcomputerinterfacesenable} discusses agent-computer interfaces for automated software engineering through iterative tool invocation. InspectCoder ~\cite{wang2025inspectcoderdynamicanalysisenabledself} enables LLM's to conduct dynamic analysis through interactive debugger control. While these approaches show remarkable results in general program repair, they mostly focus on production codebases rather than the algorithmic complexity and efficiency constraints of competitive programming.  


Although prior studies have advanced LLM-assisted debugging, they generally overlook dynamic, test-case–driven feedback. We bridge this gap by integrating brute-force generation, stress testing, and iterative refinement.

\section{Threats to Validity}
In this section, we discuss four types of threats, similar to prior research related to LLM\cite{10903889, 11028406, 10.1145/3696630.3728701}.

\noindent
\textbf{Internal Validity:}
Data contamination poses a significant threat to the internal validity of our study, as some of the problems we evaluate may already exist in the training data of the LLM.

\noindent
\textbf{External Validity:}
In this work, we use a dataset of 100 code submissions. Although the dataset is relatively small, we ensured generalizability by collecting code from 10 different users for each problem, resulting in 100 unique user submissions. This diversity reduces threats to external validity by incorporating variations in coding styles across different programmers. Additionally, we use the latest version of ChatGPT for our evaluation.

\noindent
\textbf{Construct Validity:}
All experimental procedures are documented, and the prompts for data collection and solution evaluation are publicly available to promote reproducibility and transparency.

\noindent
\textbf{Conclusion Validity:}
To reduce bias, all evaluations are conducted using a consistent prompt. Furthermore, manual data annotations are validated through independent review by two separate programmers.

\section{Discussion \& Future Scope}
Our study shows that LLMs can effectively assist in debugging, but their performance relies on execution feedback.

While we focus on competitive programming, \Name{} has broader applications in software engineering. Real-world systems rarely provide failing test cases, requiring developers to manually identify edge cases. Automating test case generation through stress testing can reveal these cases, and \Name{} demonstrates how combining generated test cases with LLM-based debugging can support software maintenance, program repair, and automated debugging pipelines. A key limitation is its dependence on a correct brute-force solution and effective test case generation, which may not scale to large input spaces or complex problems.

In the future, \Name{} could handle more complex programming contexts by integrating runtime debuggers and symbolic execution tools, such as GDB ~\cite{gdb_manual} and KLEE ~\cite{cadar2008klee}, providing richer execution feedback and exposing hidden edge cases. Adaptive prompting could reformulate failed debugging attempts with new constraints. \Name{} could also integrate into educational platforms as a teaching assistant or into professional IDEs and version control systems, offering real-time analysis, fix suggestions, and automated refactoring. These extensions would expand \Name{}’s impact across research, learning, and professional programming.
\ignore{
\section{Discussion} 
Our study shows that LLMs can effectively assist in debugging tasks, but their performance depends on the availability of execution feedback. 
\ignore{LLMs often succeed in zero-shot debugging for simple or pattern-based errors, but they struggle in cases involving deeper logical flaws or edge cases where no failure context is provided. This limitation results in redundant debugging attempts.

A key observation from our manual study is that providing failing test cases significantly improves debugging performance in complex problems. Failed test cases help LLMs localize errors and refine their reasoning. This observation motivates \Name{}, which introduces an iterative, feedback-driven debugging workflow where LLMs improve code based on failure points rather than static prompts.}
Although our evaluation focuses on competitive programming, \Name{} has broader implications for software engineering. Real-world systems rarely provide failing test cases, so developers must manually identify edge cases to refine program logic. Automating test case generation through stress testing can reveal such cases, and \Name{} shows how combining generated test cases with LLM-based debugging can support software maintenance, program repair, and automated debugging pipelines.

A key limitation is that \Name{} depends on a correct brute-force solution and effective test case generation, which may not scale to large input spaces or problems with complex constraints.
}

\section{Conclusion}
In this paper, we conduct an initial manual and empirical study on LLMs’ performance in debugging competitive problems using zero-shot prompting. Based on this study, we introduce \Name{}, a novel debugging approach that combines the capabilities of Large Language Models (LLMs) with an iterative, test-case–driven method. \Name{} integrates an iterative feedback loop guided by failing test cases and enhances the efficiency and effectiveness of debugging. Our study shows that this approach expands debugging techniques and simplifies the process compared to manual methods. \Name{} holds strong potential to serve as a debugging assistant tool for developers.

\section{Acknowledgment}
This work was supported in part by NSF grants CCF-2348277 and CCF-2518445 \cite{2024nsf....2348277Z}.


\bibliographystyle{ACM-Reference-Format}
\bibliography{sample-base}

@String{Computing = "Computing" }

@String{Computer = "{IEEE} Computer" }

@MISC{def:debug,
    author ={Margaret Rouse},
    title = {Debugging},
    howpublished = {http://searchsoftwarequality.techtarget.com/ definition/debugging},
    year = {2017, Dec 12}
}

@BOOK{Steve:Code,
  TITLE = "Code Complete",
  SUBTITLE = "",
  AUTHOR = "Steve McConnell",
  year = "2004", 
  PUBLISHER = "Microsoft Press"
}

@article{Lu:2008:LMC:1353535.1346323,
 author = {Lu, Shan and Park, Soyeon and Seo, Eunsoo and Zhou, Yuanyuan},
 title = {Learning from Mistakes: A Comprehensive Study on Real World Concurrency Bug Characteristics},
 journal = {SIGOPS Oper. Syst. Rev.},
 issue_date = {March 2008},
 volume = {42},
 number = {2},
 month = mar,
 year = {2008},
 issn = {0163-5980},
 pages = {329--339},
 numpages = {11},
 url = {http://doi.acm.org.ezproxy.uky.edu/10.1145/1353535.1346323},
 doi = {10.1145/1353535.1346323},
 acmid = {1346323},
 publisher = {ACM},
 address = {New York, NY, USA},
}

@inproceedings{cadar2008klee,
  title={Klee: unassisted and automatic generation of high-coverage tests for complex systems programs.},
  author={Cadar, Cristian and Dunbar, Daniel and Engler, Dawson R and others},
  booktitle={OSDI},
  volume={8},
  pages={209--224},
  year={2008}
}

@manual{gdb_manual,
  author = {{GNU Project}},
  title = {Debugging with GDB},
  year = {2026},
  url = {https://www.gnu.org/software/gdb/documentation/},
  note = {Accessed: 2026-03-19}
}

@INPROCEEDINGS{10903889,
  author={Pyreddy, Shireesh Reddy and Zaman, Tarannum Shaila},
  booktitle={2025 IEEE 15th Annual Computing and Communication Workshop and Conference (CCWC)}, 
  title={EmoXpt: Analyzing Emotional Variances in Human Comments and LLM-Generated Responses}, 
  year={2025},
  volume={},
  number={},
  pages={00088-00094},
  doi={10.1109/CCWC62904.2025.10903889}}

@article{lota5060080recent,
  title={Recent Trends and Challenges in Using Nlp Techniques in Software Debugging: A Systematic Literature Review},
  author={Lota, Lutfun Nahar and Zaman, Tarannum Shaila and Azwad, Mirza Mohammad and Farah, Labiba and Chowdhury, Abrar and Anjum, Zaarin and Islam, Chadni and Kamal, Abu Raihan Mostofa},
  journal={Available at SSRN 5060080}
}

@misc{chen2021evaluatinglargelanguagemodels,
      title={Evaluating Large Language Models Trained on Code}, 
      year={2021},
      eprint={2107.03374},
      archivePrefix={arXiv},
      primaryClass={cs.LG},
      url={https://arxiv.org/abs/2107.03374}, 
}

@INPROCEEDINGS{11028406,
  author={Hossain, Md Sifat and Tabassum, Anika and Arefin, Md. Fahim and Shaila Zaman, Tarannum},
  booktitle={2025 IEEE/ACM International Workshop on Large Language Models for Code (LLM4Code)}, 
  title={LLM-ProS: Analyzing Large Language Models’ Performance in Competitive Problem Solving}, 
  year={2025},
  volume={},
  number={},
  pages={80-87},
  doi={10.1109/LLM4Code66737.2025.00015}}

@misc{codef,
      title={Codeforces}, 
      url={https://codeforces.com/}, 
}

@misc{gpt5,
      title={GPT-5 is here}, 
      url={https://openai.com/gpt-5/}, 
}

@article{10.1145/3729355,
author = {Levin, Kyla H. and van Kempen, Nicolas and Berger, Emery D. and Freund, Stephen N.},
title = {ChatDBG: Augmenting Debugging with Large Language Models},
year = {2025},
issue_date = {July 2025},
publisher = {Association for Computing Machinery},
address = {New York, NY, USA},
volume = {2},
number = {FSE},
url = {https://doi.org/10.1145/3729355},
doi = {10.1145/3729355},
journal = {Proc. ACM Softw. Eng.},
month = jun,
articleno = {FSE085},
numpages = {22}
}

@misc{kang2023explainableautomateddebugginglarge,
      title={Explainable Automated Debugging via Large Language Model-driven Scientific Debugging}, 
      author={Sungmin Kang and Bei Chen and Shin Yoo and Jian-Guang Lou},
      year={2023},
      eprint={2304.02195},
      archivePrefix={arXiv},
      primaryClass={cs.SE},
      url={https://arxiv.org/abs/2304.02195}, 
}

@misc{ashrafi2025enhancingllmcodegeneration,
      title={Enhancing LLM Code Generation: A Systematic Evaluation of Multi-Agent Collaboration and Runtime Debugging for Improved Accuracy, Reliability, and Latency}, 
      author={Nazmus Ashrafi and Salah Bouktif and Mohammed Mediani},
      year={2025},
      eprint={2505.02133},
      archivePrefix={arXiv},
      primaryClass={cs.SE},
      url={https://arxiv.org/abs/2505.02133}, 
}

@misc{lee2024unifieddebuggingapproachllmbased,
      title={A Unified Debugging Approach via LLM-Based Multi-Agent Synergy}, 
      author={Cheryl Lee and Chunqiu Steven Xia and Longji Yang and Jen-tse Huang and Zhouruixin Zhu and Lingming Zhang and Michael R. Lyu},
      year={2024},
      eprint={2404.17153},
      archivePrefix={arXiv},
      primaryClass={cs.SE},
      url={https://arxiv.org/abs/2404.17153}, 
}

@article{tian2024debugbench,
  title={Debugbench: Evaluating debugging capability of large language models},
  author={Tian, Runchu and Ye, Yining and Qin, Yujia and Cong, Xin and Lin, Yankai and Pan, Yinxu and Wu, Yesai and Hui, Haotian and Liu, Weichuan and Liu, Zhiyuan and others},
  journal={arXiv preprint arXiv:2401.04621},
  year={2024}
}

@inproceedings{yang2025coast,
  title={COAST: Enhancing the Code Debugging Ability of LLMs through Communicative Agent Based Data Synthesis},
  author={Yang, Weiqing and Wang, Hanbin and Liu, Zhenghao and Li, Xinze and Yan, Yukun and Wang, Shuo and Gu, Yu and Yu, Minghe and Liu, Zhiyuan and Yu, Ge},
  booktitle={Findings of the Association for Computational Linguistics: NAACL 2025},
  pages={2570--2585},
  year={2025}
}

@article{jiang2024ledex,
  title={Ledex: Training LLMs to better self-debug and explain code},
  author={Jiang, Nan and Li, Xiaopeng and Wang, Shiqi and Zhou, Qiang and Hossain, Soneya B and Ray, Baishakhi and Kumar, Varun and Ma, Xiaofei and Deoras, Anoop},
  journal={Advances in Neural Information Processing Systems},
  volume={37},
  pages={35517--35543},
  year={2024}
}

@inproceedings{jin2024rgd,
  title={Rgd: Multi-llm based agent debugger via refinement and generation guidance},
  author={Jin, Haolin and Sun, Zechao and Chen, Huaming},
  booktitle={2024 IEEE International Conference on Agents (ICA)},
  pages={136--141},
  year={2024},
  organization={IEEE}
}

@inproceedings{tsai2024rtlfixer,
  title        = {RTLFixer: Automatically Fixing RTL Syntax Errors with Large Language Models},
  author       = {Tsai, YunDa and Liu, Mingjie and Ren, Haoxing},
  booktitle    = {Proceedings of the 61st ACM/IEEE Design Automation Conference (DAC)},
  pages        = {1--6},
  year         = {2024},
  publisher    = {ACM/IEEE},
  doi          = {10.1145/3649329.3656906}
}

@inproceedings{epperson2025agdebugger,
  title        = {Interactive Debugging and Steering of Multi-Agent AI Systems},
  author       = {Epperson, Will and Fang, Amanda and Zhang, Kaixuan and Amershi, Saleema and Weld, Daniel S. and Kamar, Ece},
  booktitle    = {Proceedings of the 2025 CHI Conference on Human Factors in Computing Systems},
  pages        = {1--13},
  year         = {2025},
  publisher    = {ACM},
  doi          = {10.1145/3613904.3642700}
}

@inproceedings{10.1145/3696630.3728701,
author = {Al Hasan, Alif and Saha, Subarna and Imran, Mia Mohammad and Zaman, Tarannum Shaila},
title = {LLPut: Investigating Large Language Models for Bug Report-Based Input Generation},
year = {2025},
isbn = {9798400712760},
publisher = {Association for Computing Machinery},
address = {New York, NY, USA},
url = {https://doi.org/10.1145/3696630.3728701},
doi = {10.1145/3696630.3728701},
booktitle = {Proceedings of the 33rd ACM International Conference on the Foundations of Software Engineering},
pages = {1652–1659},
numpages = {8},
location = {Clarion Hotel Trondheim, Trondheim, Norway},
series = {FSE Companion '25}
}

@misc{Batch,
      title={Batch Script Tutorial}, 
      url={https://www.tutorialspoint.com/batch_script/index.htm}
}

@MISC{2024nsf....2348277Z,
       author = {{Zaman}, Tarannum Shaila},
        title = "{CRII: SHF: An Automated and User-centered Framework for Reproducing System-level Concurrency Bugs by Analyzing Bug Reports}",
 howpublished = {NSF Award Number 2348277. Directorate for Computer and Information Science and Engineering, Division of Computing and Communication Foundations. 2024.},
         year = 2024,
        month = jun,
        pages = {48277},
       adsurl = {https://ui.adsabs.harvard.edu/abs/2024nsf....2348277Z}
}

@misc{wang2025inspectcoderdynamicanalysisenabledself,
      title={InspectCoder: Dynamic Analysis-Enabled Self Repair through interactive LLM-Debugger Collaboration}, 
      author={Yunkun Wang and Yue Zhang and Guochang Li and Chen Zhi and Binhua Li and Fei Huang and Yongbin Li and Shuiguang Deng},
      year={2025},
      eprint={2510.18327},
      archivePrefix={arXiv},
      primaryClass={cs.SE},
      url={https://arxiv.org/abs/2510.18327}, 
}

@misc{zhang2024autocoderoverautonomousprogramimprovement,
      title={AutoCodeRover: Autonomous Program Improvement}, 
      author={Yuntong Zhang and Haifeng Ruan and Zhiyu Fan and Abhik Roychoudhury},
      year={2024},
      eprint={2404.05427},
      archivePrefix={arXiv},
      primaryClass={cs.SE},
      url={https://arxiv.org/abs/2404.05427}, 
}

@misc{yang2024sweagentagentcomputerinterfacesenable,
      title={SWE-agent: Agent-Computer Interfaces Enable Automated Software Engineering}, 
      author={John Yang and Carlos E. Jimenez and Alexander Wettig and Kilian Lieret and Shunyu Yao and Karthik Narasimhan and Ofir Press},
      year={2024},
      eprint={2405.15793},
      archivePrefix={arXiv},
      primaryClass={cs.SE},
      url={https://arxiv.org/abs/2405.15793}, 
}

@misc{copilot,
      title={Microsoft Copilot: Your AI companion}, 
      url={https://copilot.microsoft.com/}
}

@misc{Alphacode,
      title={AlphaCode Attention Visualization}, 
      url={https://alphacode.deepmind.com/}
}


\end{document}